# Relativistic Energy Analysis Of Five Dimensional *q*-Deformed Radial Rosen-Morse Potential Combined With *q*-Deformed Trigonometric Scarf Non-Central Potential Using Asymptotic Iteration Method (AIM)

S. Pramono,[1] A. Suparmi,[2] and C. Cari[2]

*Physics Department, Graduate Program, Sebelas Maret University, Jl. Ir. Sutami 36A Kentingan Surakarta 57126 Indonesia*
*Physics Department, Faculty of Mathematics and Fundamental Science, Sebelas Maret University, Jl. Ir. Sutami 36A, Kentingan Surakarta 57126 Indonesia*

**ABSTRACT**
*In this work, we study the exact solution of Dirac equation in the hyper-spherical coordinate under influence of separable q-Deformed quantum potentials. The q-deformed hyperbolic Rosen-Morse potential is perturbed by q-deformed non-central trigonometric Scarf potentials, where whole of them can be solved by using Asymptotic Iteration Method (AIM). This work is limited to spin symmetry case. The relativistic energy equation and orbital quantum number equation $l_{D-1}$ have been obtained using Asymptotic Iteration Method. The upper radial wave function equations and angular wave function equations are also obtained by using this method. The relativistic energy levels are numerically calculated using Mat Lab, the increase of radial quantum number n causes the increase of bound state relativistic energy level both in dimension D = 5 and D = 3. The bound state relativistic energy level decreases with increasing of both deformation parameter q and orbital quantum number $n_l$.*

## 1. Introduction

Dirac equation as relativistic wave equation was formulated by P.A.M Dirac in 1928, the exact solution of Dirac equation for some quantum potentials plays fundamental role in relativistic quantum mechanics.[1] In order to investigate nuclear shell model, study of spin symmetry and pseudo-spin symmetry solutions of Dirac equations have been an important field of study in nuclear physics. The concept of spin symmetry and pseudo-spin symmetry limit with nuclear shell model has been used widely in explaining a number of phenomena in nuclear physics and related field.[2] In nuclear physics, spin symmetry and pseudo-spin symmetry concepts have been used to study the aspect of deformed and super deformation nuclei. The concept of spin symmetry has been applied to the level of meson and antinucleon.[3] Pseudo-spin symmetry has been observed in deformed nuclei and can be enhanced in heavy proton-rich nuclei.[4]

Solution of Dirac equation for some potentials under limit case of spin symmetry and pseudo-spin symmetry have been investigated intensively whether in three-[5,6] two- or one-[7-13] dimensional space, and some *D*-dimensional with spherically Symmetric Spacetimes [14-15]. However, The *D*-dimensional Dirac equation with (D-1)-dimensional separable non-central potential has not been investigated yet, therefore it may be worthy to investigate Dirac equation in 5 dimensions with separable 4-dimensional non-central potential in this study.

In this recent years, some researchers have studied solution of Dirac equation with quantum potentials with different application and methods. These investigations include: Eckart potential and trigonometric Manning-Rosen potential using Asymptotic Iteration Method (AIM),[5] q-deformed hyperbolic Pöschl-Teller potential and trigonometric Scarf II non-central potential using Nikiforov-Uvarov method,[6] *q*-deformed Trigonometric Scarf potential with *q*-deformed Trigonometric Tensor Coupling potential for Spin and Pseudo-spin Symmetries Using Romanovski Polynomial,[7] generalized nuclear Wood-Saxon potential under relativistic spin symmetry limit,[8] relativistic bound states of particle in Yukawa field with Coulomb tensor interaction,[9] Rosen-Morse potential including the spin-orbit centrifugal term using Nikiforov-Uvarov (NU) method,[3] pseudo-spin symmetric solution of the Morse potential for any $\kappa$ state using AIM,[11] Scalar, Vector, and Tensor Cornell Interaction using Ansatz method,[12] Scalar and Vector generalized Isotonic Oscillators and Cornell Tensor Interaction using Ansatz method,[13] Mie-type potentials for energy dependent pseudoharmonic potential via SUSYQM,[16] trigonometric Scarf potential in *D*-dimension for spin and pseudo-spin symmetry using Nikiforov-Uvarov (NU) method,[15] Coulombic potential and its thermodynamics properties in D-dimensional space using NU method,[18] hyperbolic tangent potential and its application in material properties in *D*-dimensional space,[19] and others.

Asymptotic Iteration Method (AIM) have small deviation for determination of eigen energies and eigen functions of Dirac equation. The separable $D$-dimensional quantum potentials is not studied yet by some researchers. In this paper we use Asymptotic Iteration Method (AIM) to solve the Dirac equation under influence of separable $D$-dimensional quantum potentials. The relativistic energy levels can be obtained from calculation of relativistic energy equation using Matlab R2013a. In section 2 we present basic theory of Dirac equation in hyper-spherical coordinate with $D$-dimensional separable quantum potential. In this section also included deformed quantum potential which is proposed by Dutra in 2005.[26] In section 3 we present asymptotic iteration method. Result and disccussion are included in section 4, and in section 5 we present the special case in 3-dimensional space. in the last section we present conclusion.

## 2. Dirac equation with separable $q$-deformed quantum potetial in The Hyper-spherical Coordinates

For single particle, Dirac equation with vector potential $V(r)$ and scalar potential $S(r)$ in the hyper-spherical coordinate can be expressed as follows (in the unit $\hbar = c = 1$):[13,14]

$$\{\hat{\alpha}.\vec{p} + \hat{\beta}(M + S(\vec{r}))\}\psi(\vec{r}) = \{E - V(\vec{r})\}\psi(\vec{r}) \quad (1)$$

where $\vec{p}$, $E$, and $M$ are $D$-dimensional momentum operator, total relativistic energy and Relativistic mass of the particle respectively.

$$\hat{\alpha}_i = \begin{pmatrix} 0 & \hat{\sigma}_i \\ \hat{\sigma}_i & 0 \end{pmatrix} \quad (2)$$

$$\hat{\beta} = \begin{pmatrix} \mathbf{1} & 0 \\ 0 & -\mathbf{1} \end{pmatrix} \quad (3)$$

where $\hat{\sigma}_i$ are Pauli's matrices and $\mathbf{1}$ is the $2 \times 2$ unit matrix. Here we use relations between Pauli's matrices as

$$\hat{\sigma}_i\hat{\sigma}_j + \hat{\sigma}_j\hat{\sigma}_i = 2\delta_{ij}\mathbf{1} \quad (4)$$

The wave function of Dirac spinor can be classified in two form, upper spinor $\chi(\vec{r})$ and lower spinor $\varphi(\vec{r})$ as follows:[15,16]

$$\psi(\vec{r}) = \begin{pmatrix} \chi(\vec{r}) \\ \varphi(\vec{r}) \end{pmatrix} = \begin{pmatrix} \frac{F_{nk}(r)}{r^{\frac{D-1}{2}}} Y^{\ell}_{\ell_1,\ldots,\ell_{D-1}}(\hat{x} = \theta_1, \theta_2, \ldots, \theta_{D-1}) \\ i\frac{G_{nk}(r)}{r^{\frac{D-1}{2}}} Y^{\tilde{\ell}}_{\tilde{\ell}_1,\ldots,\tilde{\ell}_{D-1}}(\hat{x} = \theta_1, \theta_2, \ldots, \theta_{D-1}) \end{pmatrix} \quad (5)$$

By substituting Eqs. (2), (3) and (5) into Eq. (1), we get

$$\vec{\sigma}.\vec{p}\varphi(\vec{r}) = -\left(M - E + (V(\vec{r}) + S(\vec{r}))\right)\chi(\vec{r}) \quad (6)$$

$$\vec{\sigma}.\vec{p}\chi(\vec{r}) = \left(M + E - (V(\vec{r}) - S(\vec{r}))\right)\varphi(\vec{r}) \quad (7)$$

Exact spin symmetry limit is characterized with $V(\vec{r}) = S(\vec{r})$, and for spin symmetry limit, $\Delta(\vec{r}) = V(\vec{r}) - S(\vec{r})$ and $\Sigma(\vec{r}) = V(\vec{r}) + S(\vec{r})$ for pseudo-spin symmetry limit. For spin symmetry limit we have $\frac{d\Delta(\vec{r})}{dr} = 0$ and $\Delta(\vec{r}) = C_s = constant$, therefore Eqs. (6) and (7) can be rewritten as

$$\vec{\sigma}.\vec{p}\varphi(\vec{r}) = -\{(M - E) + \Sigma(r)\}\chi(\vec{r}) \quad (8)$$

$$\varphi(\vec{r}) = \frac{\vec{\sigma}.\vec{p}}{(M + E - C_s)}\chi(\vec{r}) \quad (9)$$

If Eq. (9) is subtituted into Eq. (8) we get

$$\vec{p}^2\chi(\vec{r}) = \{(M + E) - C_s\}\left[-\{(M - E) + (\Sigma(\vec{r}))\}\right]\chi(\vec{r}) \quad (10)$$

where

$$(\vec{\sigma}.\vec{p})(\vec{\sigma}.\vec{p}) = \vec{p}^2 \quad (11)$$

Using momentum operator definition in quantum mechanics, where $\vec{p} = -i\vec{\nabla}_D$. The hyper-spherical Laplacian $\vec{\nabla}^2_D$ is given as[15]

$$\nabla^2_D = \frac{1}{r^{D-1}}\frac{\partial}{\partial r}\left(r^{D-1}\frac{\partial}{\partial r}\right) + \left(\frac{L^2_{D-1}}{r^2}\right) \quad (12)$$

The eigen value of $L^2_{D-1}$ is $\ell'_{D-1}(\ell'_{D-1} + D - 2)$ and the angular momentum operator is expressed as[15]

$$L^2_{D-1} = -\left\{\frac{1}{\sin^{D-2}\theta_{D-1}}\frac{\partial}{\partial\theta_{D-1}}\left(\sin^{D-2}\theta_{D-1}\frac{\partial}{\partial\theta_{D-1}}\right) - \frac{L^2_{D-2}}{\sin^2\theta_{D-1}}\right\} \quad (13)$$

By inserting Eq.(12) into Eq. (10) we get

$$\frac{1}{r^{D-1}}\frac{\partial}{\partial r}\left(r^{D-1}\frac{\partial}{\partial r}\chi(\vec{r})\right) + \left(\frac{L^2_{D-1}}{r^2}\right)\chi(\vec{r}) - (M^2 - E^2)\chi(\vec{r}) - \{(M + E) - C_s\}\Sigma(\vec{r})\chi(\vec{r}) = 0 \quad (14)$$

The separable variable potential used in this study is $q$-deformed hyperbolic Rosen-Morse potential plus q-deformed non sentral Scarf trigonometric potential in hyper-spherical coordinate space. The effective potential can be written as:

$$\Sigma(\vec{r}) = \begin{bmatrix} \left\{-\frac{V_0}{\cosh^2_q(\alpha r)} + V_1 \tanh_q(\alpha r)\right\} + \\ \frac{1}{r^2}\left\{\frac{V(\theta_1)}{\sin^2\theta_2\sin^2\theta_3\sin^2\theta_4} + \frac{V(\theta_2)}{\sin^2\theta_3\sin^2\theta_4} \\ + \frac{V(\theta_3)}{\sin^2\theta_4} + V(\theta_4)\right\} \end{bmatrix} \quad (15)$$

with angular potentials $V(\theta_i)$ are taken as

$$V(\theta_1) = \frac{b_1^2 + a_1(a_1 - 1)}{\sin^2_q\theta_1} - \frac{2b_1(a_1 - \frac{1}{2})\cos_q\theta_1}{\sin^2_q\theta_1} \quad (16)$$

$$V(\theta_2) = \frac{b_2^2 + a_2(a_2 - 1)}{\sin^2_q\theta_2} - \frac{2b_2(a_2 - \frac{1}{2})\cos_q\theta_2}{\sin^2_q\theta_2} \quad (17)$$

$$V(\theta_3) = \frac{b_3^2 + a_3(a_3 - 1)}{\sin^2_q\theta_3} - \frac{2b_3(a_3 - \frac{1}{2})\cos_q\theta_3}{\sin^2_q\theta_3} \quad (18)$$

$$V(\theta_4) = \frac{b_4^2 + a_4(a_4 - 1)}{\sin^2_q\theta_4} - \frac{2b_4(a_4 - \frac{1}{2})\cos_q\theta_4}{\sin^2_q\theta_4} \quad (19)$$

By subtituting Eqs. (15) into Eq. (14) and using variable separation method, we get the radial part and the angular part of Dirac equations in hypersperical coordinate with $D = 5$.

### 2.1 The Radial Part

The $D$-dimensional Dirac equation with $q$-deformed hyperbolic Rosen-Morse potential plus

$q$-deformed trigonometric Scarf non-central potentials can be resolved into the form of radial part and angular part equations. The radial part of $D$-dimensional Dirac equation in this case can be expressed as

$$\frac{d^2 F_{nk}(r)}{dr^2} - \frac{\left(\ell'_{D-1} + \frac{D-1}{2}\right)\left(\ell'_{D-1} + \frac{D-3}{2}\right)}{r^2} F_{nk}(r) + \quad (20)$$

$$\left[\left\{\frac{V_0}{\cosh_q^2(\alpha r)} - V_1 \tanh_q(\alpha r)\right\}\{M+E\} - \{M+E-C_s\}\{M-E\}\right] F_{nk}(r) = 0$$

with $\lambda_4 = \lambda_{D-1} = \ell'_{D-1}\left(\ell'_{D-1} + D - 2\right)$

## 2.2 The Angular Part

The angullar part of 5-dimensional that obtained from Eqs.(14-15) can be resolved into four parts, and for $C_s = 0$, we get

$$\frac{1}{P_1(\theta_1)}\left(\frac{\partial^2 P_1(\theta_1)}{\partial \theta_1^2}\right) + \{\lambda_1 - (M+E)V(\theta_1)\} = 0 \quad (21)$$

$$\frac{1}{P_2(\theta_2)}\left\{\frac{1}{\sin\theta_2}\frac{\partial}{\partial\theta_2}\left(\sin\theta_2\frac{\partial P_2(\theta_2)}{\partial\theta_2}\right)\right\} + \quad (22)$$

$$\left\{\lambda_2 - \frac{\lambda_1}{\sin^2\theta_2} - (M+E)V(\theta_2)\right\} = 0$$

$$\frac{1}{P_3(\theta_3)}\frac{1}{\sin^2\theta_3}\frac{\partial}{\partial\theta_3}\left(\sin^2\theta_3\frac{\partial P_3(\theta_3)}{\partial\theta_3}\right) + \quad (23)$$

$$\left\{\lambda_3 - \frac{\lambda_2}{\sin^2\theta_3} - (M+E)V(\theta_3)\right\} = 0$$

$$\frac{1}{P_4(\theta_4)}\frac{1}{\sin^3\theta_4}\frac{\partial}{\partial\theta_4}\left(\sin^3\theta_4\frac{\partial P_4(\theta_4)}{\partial\theta_4}\right) + \quad (24)$$

$$\left\{\lambda_4 - \frac{\lambda_3}{\sin^2\theta_4} - \{M+E\}V(\theta_4)\right\} = 0$$

Eqs. (21-24) are angular part of Dirac equation for $\theta_1$ until $\theta_4$ respectively. The $D$-dimensional relativistic wave functions and orbital quantum numbers are obtained from those equations.

## 3. Review of Asymptotic Iteration Method (AIM)

Asymptotic Iteration Methods (AIM) is an alternative method which have accuraccy and high efficiency to determine eigen energies and eigen functions for analytically solvable hyperbolic like potential. Asymptotic Iteration method is also giving solution for exactly solvable problem.[18]

AIM is used to solve The second orde homogen linear Eq. as follows:[5, 21-24]

$$y_n''(x) = \lambda_0(x) y_n'(x) + S_0(x) y_n(x) \quad (25)$$

where $\lambda_0(x) \neq 0$ and prime symbol refers to derivation along $x$. The others parameter $n$ is interpreted as radial quantum number. Variabel $\lambda_0(x)$ and $S_0(x)$ is a variabel that can be differentiated along $x$. To get the solution of Eq. (25), we have to differentiate Eq. (25) along $x$, and then we get

$$y_n'''(x) = \lambda_1(x) y_n'(x) + S_1(x) y_n(x) \quad (26)$$

where

$$\lambda_1(x) = \lambda_0'(x) + S_0(x) + \lambda_0^2(x) \quad (27)$$

$$S_1(x) = S_0'(x) + S_0(x)\lambda_0(x) \quad (28)$$

Asymptotic Iteration Method (AIM) and can be applied exactly in the different problem if the wave function have been known and fullfill boundary condition zero (0) and infinity ($\infty$).

Eq. (2) can be simply iterated until $(k+1)$ and $(k+2)$, $k = 1, 2, 3, ...$ and then we get

$$y_n^{k+1}(x) = \lambda_{k-1}(x) y_n'(x) + S_{k-1}(x) y_n(x) \quad (29)$$

$$y_n^{k+2}(x) = \lambda_k(x) y_n'(x) + S_k(x) y_n(x) \quad (30)$$

$$\lambda_k(x) = \lambda_{k-1}'(x) + S_{k-1}(x) + \lambda_0(x)\lambda_{k-1}(x) \quad (31)$$

$$S_k(x) = S_{k-1}'(x) + S_0(x)\lambda_{k-1}(x) \quad (32)$$

which is called as recurrence relation. Eigen value can be found using equation given as

$$\Delta_k(z) = \lambda_k(z) s_{k-1}(z) - \lambda_{k-1}(z) s_k(z) = 0 \quad (33)$$

where $k = 1,2,3, ...$ is the iteration number and $n_r$ is representation of radial quantum number.

Eq.(2) is the second orde homogen linear Equation which can be solved by comparing it with the second orde linear equation as follow:[21]

$$y''(z) = 2\left(\frac{az^{N+1}}{1-bz^{N+2}} - \frac{t+1}{z}\right) y'(z) - \frac{wz^N}{1-bz^{N+2}} y(z) \quad (34)$$

where

$$(\sigma)_n = \frac{\Gamma(\sigma+n)}{\Gamma(\sigma)}, \quad \sigma = \frac{2t+N+3}{N+2} \quad (35)$$

and

$$\rho = \frac{(2t+1)b + 2a}{(N+2)b} \quad (36)$$

The wave function of Eq. (2) is the solution of Eq.(34) which is given as [21]

$$y_n(z) = (-1)^n C_2 (N+2)^n (\sigma)_n {}_2F_1(-n, \rho+n; \sigma; bz^{N+2}) \quad (37)$$

The $q$-deformed hyperbolic and trigonometric functions are used as one of the parameters in the modified Rosen-Morse potential and non central Scarf Trigonometric potential are defined by Arai [25] some years ago as follows:

$$\sinh_q \alpha r = \frac{e^{\alpha r} - q e^{-\alpha r}}{2} \quad (38a)$$

$$\cosh_q \alpha r = \frac{e^{\alpha r} + q e^{-\alpha r}}{2} \quad (38b)$$

$$\tanh_q \alpha r = \frac{\sinh_q \alpha r}{\cosh_q \alpha r} \quad (38c)$$

$$\cosh_q^2 \alpha r - \sinh_q^2 \alpha r = q \quad (38d)$$

Deformation with $q$-parameter in the hyperbolic function can be extended into trigonometric function. Definition of trigonometric function can be arranged by the same way like in the hyperbolic function introduced by Suparmi [7] as follows :

$$\sin_q ar = \frac{e^{iar} - qe^{-iar}}{2}, \quad \cos_q ar = \frac{e^{iar} + qe^{-iar}}{2},$$
$$\cos_q^2 ar + \sin_q^2 ar = q, \quad \tan_q ar = \frac{\sin_q ar}{\cos_q ar}, \quad (39)$$
$$\sec_q ar = \frac{1}{\cos_q ar}, \quad \frac{d\sin_q ar}{dr} = a\cos_q ar,$$
$$\frac{d\cos_q ar}{dr} = -a\sin_q ar, \quad \frac{d\tan_q ar}{dr} = qa\sec_q^2 ar$$

By a convenient translation of spatial variable, one can transform the deformed potentials into the form of non-deformed potentials ones or vice-versa. In analogy to the translation of spatial variable for hyperbolic function introduced by Dutra [26], we propose the translation of spatial variable for hyperbolic and trigonometric function as follows:

$$r \to r + \frac{\ln\sqrt{q}}{\alpha}, \quad \text{and} \quad r \to r - \frac{\ln\sqrt{q}}{\alpha}, \quad (40a)$$

$$r \to r + \frac{\ln\sqrt{q}}{i\alpha}, \quad \text{and} \quad r \to r - \frac{\ln\sqrt{q}}{i\alpha} \quad (40b)$$

And then by inserting Eq. (22) into Eqs. (19) and (20) we have

$$\sinh_q \alpha r \to \sqrt{q}\sinh\alpha r; \quad \cosh_q \alpha r \to \sqrt{q}\cosh\alpha r;$$
$$\sin_q \alpha r \to \sqrt{q}\sin\alpha r; \quad \cos_q \alpha r \to \sqrt{q}\cos\alpha r; \quad (41)$$

The translation of spatial variable in Eq. (41) can be used to map the energy and wave function of non-deformed potential toward deformed potential of Scarf potential.

## 4. Result and Discussion
### 4.1 Radial Part
The radial part of Dirac equation in hyper-spherical space when $C_s=0$ can be expressed as

$$\frac{d^2 F_{nk}(r)}{dr^2} - \frac{\left(\ell'_{D-1} + \frac{D-1}{2}\right)\left(\ell'_{D-1} + \frac{D-3}{2}\right)}{r^2} F_{nk}(r) + \quad (42)$$
$$\left[\left\{\frac{V_0}{\cosh_q^2(\alpha r)} - V_1 \tanh_q(\alpha r)\right\}\{M+E\} - \{M^2 - E^2\}\right] F_{nk}(r) = 0$$

Eq. (42) can not be solved directly when $\ell'_{D-1} \neq 0$, in this condition we use approximation to solve centrifugal term with Pekeris approximation. Because we have used $q$-deformed quantum potential, the Pekeris approximation in this condition can be expressed as[3]

$$\frac{1}{r^2} = \frac{1}{r_e^2}\left(c_0 + c_1 \frac{-e^{-2\alpha r}}{1 + qe^{-2\alpha r}} + c_2 \left(\frac{-e^{-2\alpha r}}{1 + qe^{-2\alpha r}}\right)^2\right) \quad (43)$$

with

$$c_0 = 1 - \left(\frac{1 + q\exp(-2\alpha r_e)}{2\alpha r_e}\right)^2 \left(\frac{8\alpha r_e}{1 + q\exp(-2\alpha r_e)} - (3 + 2\alpha r_e)\right) \quad (44a)$$

$$c_1 = -2(\exp(2\alpha r_e) + 1)\left(3\left(\frac{1 + q\exp(-2\alpha r_e)}{2\alpha r_e}\right) - (3 + 2\alpha r_e)\left(\frac{1 + q\exp(-2\alpha r_e)}{2\alpha r_e}\right)\right) \quad (44b)$$

$$c_2 = (\exp(2\alpha r_e) + 1)^2 \left(\frac{1 + q\exp(-2\alpha r_e)}{2\alpha r_e}\right)^2 \times \quad (44c)$$
$$\left(3 + 2\alpha r_e - \frac{4\alpha r_e}{1 + q\exp(-2\alpha r_e)}\right)$$

By inserting Eq. (41) into Eq. (40) and changing from exponential form into hyperbolic form we get

$$\frac{d^2 F_{nk}(r)}{dr^2} - \left(\omega\left(c_0 - \frac{c_1}{2\sqrt{q}} + \frac{c_2}{2q}\right) + \{M^2 - E^2\}\right) F_{nk}(r) - \quad (45)$$
$$\left\{\omega\left(\frac{c_1}{2\sqrt{q}} - \frac{c_2}{2q}\right) + V_1\{M + E\}\right\} \tanh_q \alpha r F_{nk}(r) +$$
$$\left\{\frac{\omega c_2}{4} + V_0\{M + E\}\right\} \text{sech}_q^2(\alpha r) F_{nk}(r) = 0$$

with $\frac{\left(\ell'_{D-1} + \frac{D-1}{2}\right)\left(\ell'_{D-1} + \frac{D-3}{2}\right)}{r_e^2} = \omega$

where $r_e$ is equilibrium distance that can be derived by potential parameter.
Let we assume

$$\alpha^2 E' = -\left(\omega\left(c_0 - \frac{c_1}{2\sqrt{q}} + \frac{c_2}{2q}\right) + \{M^2 - E^2\}\right) \quad (46)$$

$$\alpha^2 \rho = \left\{\omega\left(\frac{c_1}{2\sqrt{q}} + \frac{c_2}{2q}\right) + V_1\{M + E\}\right\} \quad (47)$$

$$\alpha^2 \nu(\nu + 1) = \left\{\frac{\omega c_2}{4} + V_0\{M + E\}\right\} \quad (48)$$

Eqs. (46-48) are inserted into Eq. (43) so we obtain

$$\frac{d^2 F_{nk}(r)}{dr^2} - \alpha^2 \left\{\rho \tanh_q \alpha r - \nu(\nu+1)\text{sech}_q^2(\alpha r)\right\} F_{nk}(r) \quad (49)$$
$$= -\alpha^2 E' F_{nk}(r)$$

Let

$$\tanh_q \alpha r = 1 - 2z \quad (50)$$

By using Eqs. (49-50) we get

$$z(1-z)\frac{\partial^2 F_{nk}(z)}{\partial z^2} + (1-2z)\frac{\partial F_{nk}(z)}{\partial z} - \quad (51)$$
$$\frac{\left\{\rho \tanh_q \alpha r - \nu(\nu+1)\text{sech}_q^2 \alpha r\right\}}{4z(1-z)} F_{nk}(z)$$
$$= -\frac{E'}{4z(1-z)} F_{nk}(z)$$

By subtituting Eq. (50) and $\text{sech}_q^2 \alpha r = \frac{4z(1-z)}{q}$ into Eq. (51) we get

$$z(1-z)\frac{\partial^2 F_{nk}(z)}{\partial z^2} + (1-2z)\frac{\partial F_{nk}(z)}{\partial z} \quad (52)$$
$$+ \left\{\frac{\nu(\nu+1)}{q} - \left(\frac{\rho - E'}{4z}\right) - \left(\frac{-\rho - E'}{4(1-z)}\right)\right\} F_{nk}(z) = 0$$

Equation (52) has two regular singular points for $z = 0$ and $z = 1$ so the general solution from Eq. (52) is $F_{n\kappa}(z) = z^\delta (1-z)^\gamma f_n(z)$.
Let $4\delta^2 = \rho - E'$ and $4\gamma^2 = -\rho - E'$ and then subtitute it into Eq. (52) we get

$$z(1-z) f_n''(z) + \{(2\delta+1) - z(2\delta + 2\gamma + 2)\} f_n'(z) + \quad (53)$$
$$\left\{\frac{\nu(\nu+1)}{q} - (\delta + \gamma)(\delta + \gamma + 1)\right\} f_n(z) = 0$$

Eq. (53) is the second differential equation that can be manipulated into the form as in the Eq. (23)

$$f_n''(z) = \frac{\{z(2\delta+2\gamma+2)\}-(2\delta+1)}{z(1-z)} f_n'(z) + \quad (54)$$

$$\frac{\left\{(\delta+\gamma)(\delta+\gamma+1)-\frac{\nu(\nu+1)}{q}\right\}}{z(1-z)} f_n(z)$$

$$\lambda_0(z) = \frac{z(2\delta+2\gamma+2)-(2\delta+1)}{z(1-z)} \quad (55)$$

$$S_0(z) = \frac{\{(\delta+\gamma)(\delta+\gamma+1)-\frac{\nu(\nu+1)}{q}\}}{z(1-z)} \quad (56)$$

Let $A = (\delta+\gamma)(\delta+\gamma+1) - \frac{\nu(\nu+1)}{q}$ and using Eqs. (31-32), together with Eqs. (55) and (56) we get

$$\lambda_1(z) = \left\{\frac{(2\delta+1)}{z^2} + \frac{(2\gamma+1)}{(1-z)^2}\right\} + \left\{\frac{A}{z} + \frac{A}{(1-z)}\right\} + \quad (57)$$

$$\left\{-\frac{(2\delta+1)}{z} + \frac{(2\gamma+1)}{(1-z)}\right\}^2$$

$$S_1(z) = \left\{-\frac{A}{z^2} + \frac{A}{(1-z)^2}\right\} + \left\{\frac{A}{z} + \frac{A}{(1-z)}\right\} \quad (58)$$

$$\left\{-\frac{(2\delta+1)}{z} + \frac{(2\gamma+1)}{(1-z)}\right\}$$

The eigen value of Eq. (54) can be found by using Eq.(33)

$$\Delta_1 = \lambda_1 s_0 - \lambda_0 s_1 = 0 \rightarrow \varepsilon_0 = (\delta+\gamma)(\delta+\gamma+1) \quad (59)$$
$$\Delta_2 = \lambda_2 s_1 - \lambda_1 s_2 = 0 \rightarrow \varepsilon_1 = (\delta+\gamma+1)(\delta+\gamma+2)$$
$$\varepsilon_n = (\delta+\gamma+n)(\delta+\gamma+n+1)$$

where $\varepsilon_n$ is $n^{th}$ eigen value when $n = 0, 1, 2, ...$ and if $n$ is radial quantum number. By using Eqs. (31-32) and the last eqation in Eq.(59) we have

$$\varepsilon_n = \frac{\nu(\nu+1)}{q} = \left\{\frac{(\ell_4'+\frac{3}{2})(\ell_4'+\frac{1}{2})}{r_e^2 4\alpha^2 q} c_2 + \frac{V_0(M+E)}{\alpha^2 q}\right\} \quad (60)$$

The last equation in Eq. (59) can be rewritten

$$\varepsilon_n = (\delta+\gamma+(n+1))^2 - (\delta+\gamma+(n+1)) \quad (61)$$

From Eq. (61) we get relativistic energy Equation of this system is

$$(M^2-E^2)$$

$$= \alpha^2 \left[ \frac{\rho^2}{4\left(\sqrt{\varepsilon_n+\frac{1}{4}}-n-\frac{1}{2}\right)^2} + \left(\sqrt{\varepsilon_n+\frac{1}{4}}-n-\frac{1}{2}\right)^2 \right] - \omega\left(c_0 - \frac{c_1}{2\sqrt{q}} + \frac{c_2}{2q}\right) \quad (62)$$

By comparing Eq.(34) and Eq. (54) and using Eq. (37), we obtain $f_n(z)$ as

$$f_n(z) = (-1)^n C'(1)^n (2\delta+1)_n \times \quad (63)$$
$$_2F_1(-n, 2\delta+2\gamma+n+1, 2\delta+1, z)$$

From Eq. (63) we determine radial wave function for various $n$ as shown in Table 7. The unnormalized radial wave function for $D=5$, with $C'$ is normalization constant.

## 4.2 Solution of Angular Part

In this study, the four angular part of Dirac equations are presented in Eqs. (21-24), so we have to solve each equation of angular Dirac equation using AIM.

## Equation of Angular Part for $\theta_1$

We can solve Eq. (21) by changing it into the form of the second orde hypergeometric-type differential equation that is similar to Eq.(23) after we insert Eq.(16) in Eq.(21)

$$\frac{\partial^2 P_1(\theta_1)}{\partial \theta_1^2} - \{M+E\} \times$$

$$\left\{\frac{b_1^2 + a_1(a_1-1)}{\sin_q^2 \theta_1} + \frac{2b_1(a_1-\frac{1}{2})\cos_q \theta_1}{\sin_q^2 \theta_1}\right\} P_1(\theta_1) + \quad (64)$$

$$\lambda_1 P_1(\theta_1) = 0$$

and substitute $\cos_q \theta_1 = \sqrt{q}(1-2z_1)$ into Eq.(64) so we get

$$z_1(1-z_1)\frac{d^2 P_1(z_1)}{dz_1^2} + \frac{1}{2}(1-2z_1)\frac{dP_1(z_1)}{dz_1} +$$

$$\left\{ \lambda_1 - \frac{\{M+E\}\left[\frac{(b_1^2+a_1(a_1-1))}{q} + \frac{(2b_1(a_1-\frac{1}{2})\sqrt{q})}{q}\right]}{4z_1} \right. \quad (65)$$

$$\left. - \frac{\{M+E\}\left[\frac{(b_1^2+a_1(a_1-1))}{q} - \frac{(2b_1(a_1-\frac{1}{2})\sqrt{q})}{q}\right]}{4(1-z_1)} \right\} P_1(z_1) = 0$$

Eq. (66) has two regular singular points for $z_1 = 0$ and $z_1 = 1$, and then the solution of $P_1(z_1)$ is set as

$$P_1(z_1) = z_1^{\delta_{s1}}(1-z_1)^{\gamma_{s1}} f_{n_t}(z_1) \quad (66)$$

If we replace $P_1(z_1)$ in Eq. (65) with Eq. (66) and simplify it by using appropriate variabel substitution as follows

$$O_{s1}^2 = \lambda_1 \quad (67a)$$

$$2\delta_{s1}(2\delta_{s1}-1) = (M+E)\left(\frac{b_1^2+a_1(a_1-1)}{q} - \frac{2b_1(a_1-\frac{1}{2})\sqrt{q}}{q}\right) \quad (67b)$$

$$2\gamma_{s1}(2\gamma_{s1}-1) = (M+E)\left(\frac{b_1^2+a_1(a_1-1)}{q} + \frac{2b_1(a_1-\frac{1}{2})\sqrt{q}}{q}\right) \quad (67c)$$

then Eq.(65) reduces into

$$f_{n_t}''(z_1) \quad (68)$$

$$= \frac{\{(2\delta_{s1}+2\gamma_{s1}+1)z_1 - (2\delta_{s1}+\frac{1}{2})\}}{z_1(1-z_1)} f_{n_t}'(z_1) +$$

$$\frac{\{(\delta_{s1}+\gamma_{s1})^2 - O_{s1}^2\}}{z_1(1-z_1)} f_{n_t}(z_1)$$

From Eq. (69) we get

$$\lambda_0(z_1) = \frac{\{(2\delta_{s1}+2\gamma_{s1}+1)z_1 - (2\delta_{s1}+\frac{1}{2})\}}{z_1(1-z_1)}; \quad (69)$$

$$S_0(z_1) = \frac{\{(\delta_{s1} + \gamma_{s1})^2 - O_{s1}^2\}}{z_1(1-z_1)} \quad (70)$$

$$\lambda_1(z_1) = \left\{\frac{\left(2\delta_{s1}+\frac{1}{2}\right)}{z_1^2} + \frac{\left(2\gamma_{s1}+\frac{1}{2}\right)}{(1-z_1)^2}\right\} + \left\{\frac{I_{s1}}{z_1} + \frac{I_{s1}}{(1-z_1)}\right\} + \quad (71)$$

$$\left\{-\frac{\left(2\delta_{s1}+\frac{1}{2}\right)}{z_1} + \frac{\left(2\gamma_{s1}+\frac{1}{2}\right)}{(1-z_1)}\right\}^2$$

$$S_1(z_1) = \left\{-\frac{I_{s1}}{z_1^2} + \frac{I_{s1}}{(1-z_1)^2}\right\} + \left\{\frac{I_{s1}}{z_1} + \frac{I_{s1}}{(1-z_1)}\right\} \times \quad (72)$$

$$\left\{-\frac{\left(2\delta_{s1}+\frac{1}{2}\right)}{z_1} + \frac{\left(2\gamma_{s1}+\frac{1}{2}\right)}{(1-z_1)}\right\}$$

By using Eqs. (69-72) and using relation of Eq. (33) we get

$$\left(\ell_1'^2\right)_{n_l} = (\delta_{s1} + \gamma_{s1} + n_l)^2 \quad (73)$$

where $n_l$ is orbital quantum number. The angular wave function of Eq.(68) is determined by using Eq. (37) and then we get

$$f_{n_l} = (-1)^{n_l} C'(1)^{n_l} \left(2\delta_{s1} + \frac{1}{2}\right)_{n_l} \times \quad (74)$$

$${}_2F_1\left(-n_l, 2\delta_{s1} + 2\gamma_{s1} + n_l, 2\delta_{s1} + \frac{1}{2}, z\right)$$

where $C'$ is normalization constant and ${}_2F_1(a,b;c;z)$ is hypergeometry function. From Eq. (74) we solve unnormalized angular wave function as function of $z$ completely for various of $n_l$ as shown in Table 8.

### 4.2.2 Equation of Angular Part for $\theta_2, \theta_3, \theta_4$

Solution for $\theta_2$, $\theta_3$, and $\theta_4$ are determined by the same way with solution for $\theta_1$, by setting subscript $i$ with 2, 3 and 4. so we get solution for orbital quantum number for $\theta_2$, $\theta_3$, and $\theta_4$ respectively as follows:

$$(\ell_2')_{n_l} = \delta_{s2} + \gamma_{s2} + \left(n_l - \tfrac{1}{2}\right), \text{ for } \theta_2 \ n_l = 0,1,\ldots \quad (75)$$

$$(\ell_3')_{n_l} = \delta_{s3} + \gamma_{s3} + (n_l - 1), \text{ for } \theta_3 \ n_l = 0,1,\ldots \quad (76)$$

$$(\ell_4')_{n_l} = \delta_{s4} + \gamma_{s4} + \left(n_l - \tfrac{3}{2}\right), \text{ for } \theta_4 \ n_l = 0,1,\ldots \quad (77)$$

where

$$\delta_{s2} = \frac{\sqrt{\lambda_1 + (M+E)\left(\frac{b_2^2 + a_2(a_2-1)}{q} - \frac{2b_2(a_2-\frac{1}{2})}{\sqrt{q}}\right)} + \frac{1}{2}}{2} \quad (78)$$

$$\delta_{s3} = \frac{\sqrt{\lambda_2 + \frac{1}{4} + (M+E)\left(\frac{b_3^2 + a_3(a_3-1)}{q} - \frac{2b_3(a_3-\frac{1}{2})}{\sqrt{q}}\right)} + \frac{1}{2}}{2} \quad (79)$$

$$\delta_{s4} = \frac{\sqrt{\lambda_3 - \frac{1}{2} + (M+E)\left(\frac{b_4^2 + a_4(a_4-1)}{q} - \frac{2b_4(a_4-\frac{1}{2})}{\sqrt{q}}\right)} + \frac{1}{2}}{2} \quad (80)$$

$$\gamma_{s2} = \frac{\sqrt{\lambda_1 + (M+E)\left(\frac{b_2^2 + a_2(a_2-1)}{q} + \frac{2b_2(a_2-\frac{1}{2})}{\sqrt{q}}\right)} + \frac{1}{2}}{2} \quad (81)$$

$$\gamma_{s3} = \frac{\sqrt{\lambda_2 + (M+E)\left(\frac{b_3^2 + a_3(a_3-1)}{q} + \frac{2b_3(a_3-\frac{1}{2})}{\sqrt{q}}\right)} + \frac{1}{4} + \frac{1}{2}}{2} \quad (82)$$

$$\gamma_{s4} = \frac{\sqrt{\lambda_3 - \frac{1}{2} + (M+E)\left(\frac{b_4^2 + a_4(a_4-1)}{q} + \frac{2b_4(a_4-\frac{1}{2})}{\sqrt{q}}\right)} + \frac{1}{2}}{2} \quad (83)$$

The unnormalized angular wave function for $\theta_2$, $\theta_3$, and $\theta_4$ are in the same pattern with angular wave function for $\theta_1$ so we obtain the last three angular wave function by simple change of parameters in Eq. (74) with parameters in Eqs.(78-83).

The relativistic energy levels are calculated numerically using Matlab program R2013a. Table 1 shows the relativistic energy levels as a function of deformation parameter $q$, the relativistic energy $E$ decreases when deformation parameter $q$ increases. Here we apply the value of $q$ from 0.2 until 1.2 with step 0.2. The relativistic energy levels as a function of orbital quantum numbers $n_l$ are also shown in Table 1. The magnitude of Energies $E$ decrease when the orbital quantum numbers $n_l$ increase. The relativistic energy levels as a function of radial quantum number $n$ are shown in Table 2. The relativistic energies $E$ increase when radial quantum number $n$ increases. The relativistic energies numerically also change as a function of potential parameters $a_i$ and $b_i$ where $i = 1,2,3,4$ are the component of $i$-th non-central potential. The relativistic energies $E$ increase when both potential parameters in each potential component increase. This suggests that the bounded energies become less bounded with increasing of potential paremeters. The unnormalized angular wave function are listed in

Table 5. The unnnormalized radial wave functions are plotted by using Eqs. (50) and (63) are shown in the Fig. 1. From the Fig. 1 (a to c) it is seen that amplitude of the wave funtion become increases when the orbital quantum number increases. This suggests that probability to find particles is larger for higher orbital quantum number $n_l$.

**Table 1.** Relativistic Energies with Variation of $q$ and $nl$

$D = 5$, $V_0=6$, $V_1=-1$, $a_1=a_2=a_3=a_4=b_1=b_2=b_3=b_4=2$

| $q$ | $r_e(fm)$ | $n$ | $n_{l1}=n_{l2}=n_{l3}=n_{l4}$ | $E_q(fm^{-1})$ | $n_{l1}=n_{l2}=n_{l3}=n_{l4}$ | $n$ | $q$ | $r_e\,(fm)$ | $E_{n_l}(fm^{-1})$ |
|---|---|---|---|---|---|---|---|---|---|
| 0.2 | 0.0333 | 1 | 1 | -5.2311 | 0 | 1 | 1 | 0.1671 | -5.0977 |
| 0.4 | 0.0667 | 1 | 1 | -5.4766 | 1 | 1 | 1 | 0.1671 | -6.4721 |
| 0.6 | 0.1001 | 1 | 1 | -5.7371 | 2 | 1 | 1 | 0.1671 | -9.1544 |
| 0.8 | 0.1335 | 1 | 1 | -6.0240 | 3 | 1 | 1 | 0.1671 | -13.1643 |
| 1 | 0.1671 | 1 | 1 | -6.4721 | 4 | 1 | 1 | 0.1671 | -18.5283 |
| 1.2 | 0.2007 | 1 | 1 | -6.9767 | 5 | 1 | 1 | 0.1671 | -25.2837 |

**Table 2.** Relativistic Energies with Variation of radial quantum number $n$

$D = 5$, $V_0=6$, $V_1=-1$, $\alpha=0,5\;(fm^{-1})$, $r_e=0.1671\;(fm)$
$a_1=a_2=a_3=a_4=b_1=b_2=b_3=b_4=2$

| $n$ | $q$ | $n_{l1}=n_{l2}=n_{l3}=n_{l4}$ | $E_n(fm^{-1})$ |
|---|---|---|---|
| 0 | 1 | 1 | -6.4725 |
| 1 | 1 | 1 | -6.4721 |
| 2 | 1 | 1 | -6.4717 |
| 3 | 1 | 1 | -6.4713 |
| 4 | 1 | 1 | -6.4709 |

**Table 3.** The Relativistic Energies with Variation of potential parameters $a_1;a_2;a_3;a_4$ and $b_1;b_2;b_3;b_4$ dengan $M = 5; C_s = 0; n_r = 0; n_{l1} = n_{l2} = n_{l3} = n_{l4} = 0; V_0 = 6; \alpha = 0.5; q = 1; V_1 = -1; q = 1$ and $r_e = 0.1671$.

| $a_1$ | $b_1$ | $a_2$ | $b_2$ | $a_3$ | $b_3$ | $a_4$ | $b_4$ | $E_{ab}(fm)$ |
|---|---|---|---|---|---|---|---|---|
| 2 | 2 | 2 | 2 | 2 | 2 | 2 | 2 | -5.0980 |
| 4 | 4 | 2 | 2 | 2 | 2 | 2 | 2 | -5.0821 |
| 6 | 6 | 2 | 2 | 2 | 2 | 2 | 2 | -5.0561 |
| 8 | 8 | 2 | 2 | 2 | 2 | 2 | 2 | -5.0382 |
| 10 | 10 | 2 | 2 | 2 | 2 | 2 | 2 | -5.0269 |
| 2 | 2 | 4 | 4 | 2 | 2 | 2 | 2 | -5.0528 |
| 2 | 2 | 6 | 6 | 2 | 2 | 2 | 2 | -5.0319 |
| 2 | 2 | 8 | 8 | 2 | 2 | 2 | 2 | -5.0211 |
| 2 | 2 | 10 | 10 | 2 | 2 | 2 | 2 | -5.0171 |
| 2 | 2 | 2 | 2 | 4 | 4 | 2 | 2 | -5.0383 |
| 2 | 2 | 2 | 2 | 6 | 6 | 2 | 2 | -5.0136 |
| 2 | 2 | 2 | 2 | 8 | 8 | 2 | 2 | -5.0084 |
| 2 | 2 | 2 | 2 | 10 | 10 | 2 | 2 | -5.0055 |
| 2 | 2 | 2 | 2 | 2 | 2 | 4 | 4 | -5.0453 |
| 2 | 2 | 2 | 2 | 2 | 2 | 6 | 6 | -5.0266 |
| 2 | 2 | 2 | 2 | 2 | 2 | 8 | 8 | -5.0200 |
| 2 | 2 | 2 | 2 | 2 | 2 | 10 | 10 | -5.0156 |

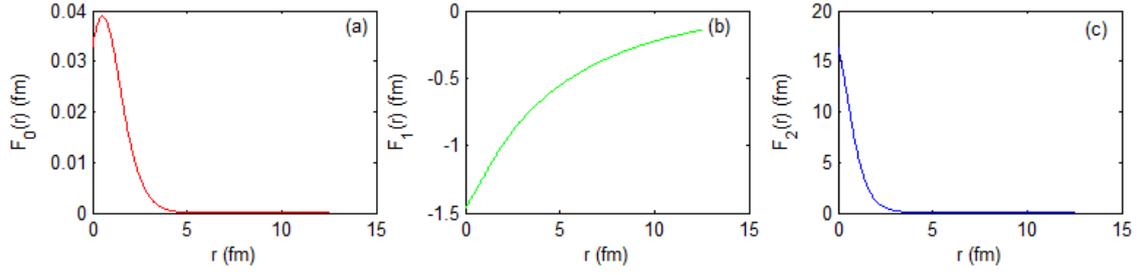

**Fig. 1** Unnormalized radial wave functions vs hyperpherical radius $r$ with various of radial quantum number, where $M = 5\,fm^{-1}, C_s = 0, V_0 = 6, V_1 = -1, r_e = 0.1671 fm$ and $n_{li} = 1, i = 1,2,3,4$. (a) for $n = 0$, (b) For $n = 1$ and (c) For $n = 2$.

**Table 5.** The Unnormalized Upper Angular Wave Function for higher dimension $D = 5$.

| $n_l$ | Unnormalized Angular Wave Function |
|---|---|
| 0 | $Q(\theta_1) = C'\left(\dfrac{1-\cos\theta_1}{2}\right)^{\delta_{s1}}\left(\dfrac{1+\cos\theta_1}{2}\right)^{\gamma_{s1}}$ |
|  | $Q(\theta_2) = C'\left(\dfrac{1-\cos\theta_2}{2}\right)^{\delta_{s2}}\left(\dfrac{1+\cos\theta_2}{2}\right)^{\gamma_{s2}}$ |
|  | $Q(\theta_3) = C'\left(\dfrac{1-\cos\theta_3}{2}\right)^{\delta_{s3}}\left(\dfrac{1+\cos\theta_3}{2}\right)^{\gamma_{s3}}$ |
|  | $Q(\theta_4) = C'\left(\dfrac{1-\cos\theta_4}{2}\right)^{\delta_{s4}}\left(\dfrac{1+\cos\theta_4}{2}\right)^{\gamma_{s4}}$ |
| 1 | $Q(\theta_1) = C'\left(\dfrac{1-\cos\theta_1}{2}\right)^{\delta_{s1}}\left(\dfrac{1+\cos\theta_1}{2}\right)^{\gamma_{s1}}(-1)\left(2\delta_{s1}+\dfrac{1}{2}\right)\left[1+\dfrac{(-1)(2\delta_{s1}+2\gamma_{s1}+1)\left(\dfrac{1-\cos\theta_1}{2}\right)}{\left(2\delta_{s1}+\dfrac{1}{2}\right)}\right]$ |
|  | $Q(\theta_2) = C'\left(\dfrac{1-\cos\theta_2}{2}\right)^{\delta_{s2}}\left(\dfrac{1+\cos\theta_2}{2}\right)^{\gamma_{s2}}(-1)\left(2\delta_{s2}+\dfrac{1}{2}\right)\left[1+\dfrac{(-1)(2\delta_{s2}+2\gamma_{s2}+1)\left(\dfrac{1-\cos\theta_2}{2}\right)}{\left(2\delta_{s2}+\dfrac{1}{2}\right)}\right]$ |
|  | $Q(\theta_3) = C'\left(\dfrac{1-\cos\theta_3}{2}\right)^{\delta_{s3}}\left(\dfrac{1+\cos\theta_3}{2}\right)^{\gamma_{s3}}(-1)\left(2\delta_{s3}+\dfrac{1}{2}\right)\left[1+\dfrac{(-1)(2\delta_{s3}+2\gamma_{s3}+1)\left(\dfrac{1-\cos\theta_3}{2}\right)}{\left(2\delta_{s3}+\dfrac{1}{2}\right)}\right]$ |
|  | $Q(\theta_4) = C'\left(\dfrac{1-\cos\theta_4}{2}\right)^{\delta_{s4}}\left(\dfrac{1+\cos\theta_4}{2}\right)^{\gamma_{s4}}(-1)\left(2\delta_{s4}+\dfrac{1}{2}\right)\left[1+\dfrac{(-1)(2\delta_{s4}+2\gamma_{s4}+1)\left(\dfrac{1-\cos\theta_4}{2}\right)}{\left(2\delta_{s4}+\dfrac{1}{2}\right)}\right]$ |

## 5 Special Case For Three Dimensional System

In three dimensional case, D=3, the bound state relativistic energy levels are calculated numerically from Eqs.(65) and (71a-71c, 76, 78-86) using matlab software R2013a and are presented at Table 6, 7 and 8. By using Eqs. (50) and (66) the unnormalized radial wave function are listed in Table 9. In the Table 6, the relativistic energy levels decrease with the increase of both deformation parameter $q$ and orbital quantum number $n_l$, where $n_{l1}$ is orbital quantum number for $\theta_1$ and $n_{l2}$ is orbital quantum number for $\theta_2$. From Table 7 we can conclude that the value of relativistic energies increase with the increase of the radial quantum number $n$ and either with the presence/absence of angular potential parameter. The angular potential parameters $a$ and $b$ influence to the relativistic energy level, where the relativistic energy levels increase by increasing of the angular potential parameter. The unnormalized upper radial wave functions are listed in the Table 9 when the radial quantum numbers are $n = 0$ and $n = 1$ and $C'$ is the normalization constant. The unnormalized angular wave functions as a functions of $\theta_1$ and $\theta_2$ are listed in Table 10. Without the presence of the non-central potential, degeneracy energy spectra occurs in the spin doublets, with quantum numbers $(n,\ell, j = \ell + 1/2)$ and $(n,\ell, j = \ell - 1/2)$, where $n$, $\ell$ and $j$ are the radial, the orbital and the total angular momentum quantum numbers respectively, for example in $(np_{1/2}, np_{3/2})$ for $\ell = 1$ but for difference values of $j = 1/2$ and $3/2$, $(nd_{3/2}, nd_{5/2})$ for $\ell = 2$ with $(j = 3/2$ and $5/2)$, $(nf_{5/2}, nf_{7/2})$ for $\ell = 3$ with $(j = 5/2$ and 7/2), and ($ng_{7/2}$, $ng_{9/2}$) for $\ell = 4$ with ($j = 7/2$ and 9/2). The degeneracy energies can be removed with the presence of non-central potential, by changing $\ell \to K$, where $K = -\ell - 1$ and $K = \ell$ for $K < 0$ and $K > 0$.

**Table 6.** Relativistic Energy with Variation of $q$ and $nl$

$V_0$=6, $V_1$=-1, α=0,5 ($fm^{-1}$) $a_1$=$a_2$=$b_1$=$b_2$=2

| q | $r_e$ (fm) | n | $n_{l1}$=$n_{l2}$ | $E_q(fm^{-1})$ | $n_{l1}$=$n_{l2}$ | nr | q | $r_e$(fm) | $E_{n_l}(fm^{-1})$ |
|---|---|---|---|---|---|---|---|---|---|
| 0.2 | 0.0333 | 1 | 1 | -5.1360 | 0 | 1 | 1 | 0.1671 | -5.1394 |
| 0.4 | 0.0667 | 1 | 1 | -5.2867 | 1 | 1 | 1 | 0.1671 | -6.1034 |
| 0.6 | 0.1001 | 1 | 1 | -5.4552 | 2 | 1 | 1 | 0.1671 | -7.9374 |
| 0.8 | 0.1335 | 1 | 1 | -5.6616 | 3 | 1 | 1 | 0.1671 | -10.6614 |
| 1 | 0.1671 | 1 | 1 | -6.1034 | 4 | 1 | 1 | 0.1671 | -14.3053 |
| 1.2 | 0.2007 | 1 | 1 | -6.5579 | | | | | |

**Table 7.** Relativistic Energies with Variation of $nr$ and potential parameter

$M$=5, $V_0$=6, $V_1$=-1, $q$=1, α=0,5 ($fm^{-1}$), $r_e$ = 0.1671($fm$)

| n | $n_{l1}$=$n_{l2}$ | $a_1$=$a_2$=$b_1$=$b_2$ | $E_{nn_{l1}n_{l2}}(fm^{-1})$ | n | $n_{l1}$=$n_{l2}$ | $a_1$=$a_2$=$b_1$=$b_2$ | $E_{nn_{l1}n_{l2}}(fm^{-1})$ |
|---|---|---|---|---|---|---|---|
| 0 | 1 | 2 | -6.1042 | 0 | 1 | 0 | -0.0616 |
| 1 | 1 | 2 | -6.1034 | 1 | 1 | 0 | -0.0564 |
| 2 | 1 | 2 | -6.1024 | 2 | 1 | 0 | -0.0509 |
| 3 | 1 | 2 | -6.1014 | 3 | 1 | 0 | -0.0454 |
| 4 | 1 | 2 | -6.1004 | 4 | 1 | 0 | -0.0398 |

**Table 8.** Relativistic Energies for $M = 1fm^{-1}$, $C_s = 0$, $V_0 = 6$, $V_1 = -1$ without Presence of q-Deformed Non-Central Potentials.

| $\ell$ | n | K | ($\ell,j$) | $E_{nK}^{ss}$ (fm$^{-1}$), $q$=1, $r_e$=0.1671 | $E_{nK}^{ss}$ (fm$^{-1}$), $q$=1.2, $r_e$=0.2007 | $E_{nK}^{ss}$ (fm$^{-1}$), $q$=1.4, $r_e$=0.2334 |
|---|---|---|---|---|---|---|
| 1 | 0 | -2 | 0p$_{3/2}$ | -0.5028 | -0.5014 | -0.5006 |
| 2 | 0 | -3 | 0d$_{5/2}$ | -0.5008 | -0.5002 | -0.4999 |
| 3 | 0 | -4 | 0f$_{7/2}$ | -0.5002 | -0.4999 | -0.4996 |
| 4 | 0 | -5 | 0g$_{9/2}$ | -0.5000 | -0.4998 | -0.4995 |
| 1 | 1 | -2 | 1p$_{3/2}$ | -0.5042 | -0.5028 | -0.5020 |
| 2 | 1 | -3 | 1d$_{5/2}$ | -0.5013 | -0.5007 | -0.5003 |
| 3 | 1 | -4 | 1f$_{7/2}$ | -0.5005 | -0.5002 | -0.4999 |
| 4 | 1 | -5 | 1g$_{9/2}$ | -0.5002 | -0.4999 | -0.4997 |
| 1 | 0 | 1 | 0p$_{1/2}$ | -0.5028 | -0.5014 | -0.5006 |
| 2 | 0 | 2 | 0d$_{3/2}$ | -0.5008 | -0.5002 | -0.4999 |
| 3 | 0 | 3 | 0f$_{5/2}$ | -0.5002 | -0.4999 | -0.4996 |
| 4 | 0 | 4 | 0g$_{7/2}$ | -0.5000 | -0.4998 | -0.4995 |
| 1 | 1 | 1 | 1p$_{1/2}$ | -0.5042 | -0.5028 | -0.5020 |
| 2 | 1 | 2 | 1d$_{3/2}$ | -0.5013 | -0.5007 | -0.5003 |
| 3 | 1 | 3 | 1f$_{5/2}$ | -0.5005 | -0.5002 | -0.4999 |
| 4 | 1 | 4 | 1g$_{7/2}$ | -0.5002 | -0.4999 | -0.4997 |

**Table 9.** Unnormalized Upper Radial Wave Function for dimension $D = 3$

| N | Radial Wave Function |
|---|---|
| 0 | $F_0(r) = C' \left( \dfrac{1 - \tanh_q \alpha r}{2} \right)^\delta \left( \dfrac{1 + \tanh_q \alpha r}{2} \right)^\gamma$ |
| 1 | $F_1(r) = C' \left( \dfrac{1 - \tanh_q \alpha r}{2} \right)^\delta \left( \dfrac{1 + \tanh_q \alpha r}{2} \right)^\gamma (-1) C'(2\delta+1) \left[ 1 + \dfrac{(-1)_1 (2\delta + 2\gamma + 2)_1 \left( \dfrac{1 - \tanh_q \alpha r}{2} \right)^1}{(2\delta+1)_1 \, 1!} \right]$ |

Table 10. The Unnormalized Upper Anguler Wave Function for dimension $D = 3$.

| $n_l$ | Unnormalized Angular Wave Function |
|---|---|
| 0 | $Q(\theta_1) = C'\left(\dfrac{1-\cos\theta_1}{2}\right)^{\delta_{s1}}\left(\dfrac{1+\cos\theta_1}{2}\right)^{\gamma_{s1}}$ |
|   | $Q(\theta_2) = C'\left(\dfrac{1-\cos\theta_2}{2}\right)^{\delta_{s2}}\left(\dfrac{1+\cos\theta_2}{2}\right)^{\gamma_{s2}}$ |
| 1 | $Q(\theta_1) = C'\left(\dfrac{1-\cos\theta_1}{2}\right)^{\delta_{s1}}\left(\dfrac{1+\cos\theta_1}{2}\right)^{\gamma_{s1}}(-1)\left(2\delta_{s1}+\dfrac{1}{2}\right)\left[1+\dfrac{(-1)(2\delta_{s1}+2\gamma_{s1}+1)\left(\dfrac{1-\cos\theta_1}{2}\right)}{\left(2\delta_{s1}+\dfrac{1}{2}\right)}\right]$ |
|   | $Q(\theta_2) = C'\left(\dfrac{1-\cos\theta_2}{2}\right)^{\delta_{s2}}\left(\dfrac{1+\cos\theta_2}{2}\right)^{\gamma_{s2}}(-1)\left(2\delta_{s2}+\dfrac{1}{2}\right)\left[1+\dfrac{(-1)(2\delta_{s2}+2\gamma_{s2}+1)\left(\dfrac{1-\cos\theta_2}{2}\right)}{\left(2\delta_{s2}+\dfrac{1}{2}\right)}\right]$ |

# 6 Conclusion

In this study, we have obtained the bound state solution of the 5 dimensional Dirac equation with separable $q$-deformed quantum potential under condition of spin symmetry. The uppper component of Dirac spinors and relativistic energy have been obtained using Asymptotic Iteration Method (AIM). The numerical result shows that the bound state relativistic energy level for spin symmetry case with both dimension $D = 5$ and $D = 3$ increase with increasing of radial quantum number $n$, and decrease with increasing of both deformation parameter $q$ and orbital quantum number $n_l$. The degeneracy energy states occurs in the absence of q-deformed non-central potential.

# 7 Acknowledgments

This Research is partly supported by Higher Education Project Grant with contract no. 632/UN27/27.21/LT/2016.